\documentstyle[12pt,epsf]{article}

\def\thefootnote{\fnsymbol{footnote}}

\begin{document}

\newcommand{\gsim}{ \mathop{}_{\textstyle \sim}^{\textstyle >} }
\newcommand{\lsim}{ \mathop{}_{\textstyle \sim}^{\textstyle <} }
\newcommand{\vev}[1]{ \left\langle {#1} \right\rangle }
\newcommand{\lsp}{ \left ( }
\newcommand{\rsp}{ \right ) }
\newcommand{\lmp}{ \left \{ }
\newcommand{\rmp}{ \right \} }
\newcommand{\llp}{ \left [ }
\newcommand{\rlp}{ \right ] }
\newcommand{\labs}{ \left | }
\newcommand{\rabs}{ \right | }

\newcommand{\K}{ {\rm K} }
\newcommand{\EV}{ {\rm eV} }
\newcommand{\KEV}{ {\rm keV} }
\newcommand{\MEV}{ {\rm MeV} }
\newcommand{\GEV}{ {\rm GeV} }
\newcommand{\TEV}{ {\rm TeV} }

\begin{titlepage}
\begin{center}

\hfill    ICRR-Report-340-95-6\\
\hfill    LBL-37715 \\
\hfill    UT-719\\
\hfill    hep-ph/???????\\

\vskip .5in

{\Large \bf Constraint on the Reheating Temperature from 
the Decay of the Polonyi Field}
\footnote{This work was supported by the Director, Office of Energy 
Research, Office of High Energy and Nuclear Physics, Division of High 
Energy Physics of the U.S. Department of Energy under Contract 
DE-AC03-76SF00098.}
\vskip .50in

{\large M.~Kawasaki$^a$, T.~Moroi$^b$ and T.~Yanagida$^{c,d}$}

\vskip .5in

$^a${\it Institute for Cosmic Ray Research, University of Tokyo, Tokyo 
188, JAPAN}

$^b${\it Theoretical Physics Group, Lawrence Berkeley Laboratory, 
University of California, Berkeley, CA 94720, U.S.A.}

$^c${\it Department of Physics, University of Tokyo, Tokyo 133, JAPAN}

$^d${\it Institute for Theoretical Physics, University of California,
Santa Barbara, CA 93106, U.S.A.}

\end{center}

\vskip .5in

\begin{abstract}

    We study the Polonyi problem in the framework of no-scale type
    supergravity models. We show that the lightest superparticle (LSP)
    produced in the decay of the Polonyi field may contribute too much
    to the present density of the universe. By requiring that LSP
    should not overclose the universe, we obtain a stringent
    constraint on the reheating temperature after the decay of the
    Polonyi field. We calculate the LSP density with physical
    parameters obtained by solving renormalization group equations in
    the minimal supersymmetric SU(5) model and find that the reheating
    temperature should be greater than about 100MeV which corresponds
    to $O$(100)TeV of the Polonyi mass.

\end{abstract}
\end{titlepage}
\renewcommand{\thepage}{\roman{page}}
\setcounter{page}{2}
\mbox{ }

\vskip 1in

\begin{center}
{\bf Disclaimer}
\end{center}

\vskip .2in

\begin{scriptsize}
\begin{quotation}
This document was prepared as an account of work sponsored by the United
States Government. While this document is believed to contain correct 
information, neither the United States Government nor any agency
thereof, nor The Regents of the University of California, nor any of their
employees, makes any warranty, express or implied, or assumes any legal
liability or responsibility for the accuracy, completeness, or usefulness
of any information, apparatus, product, or process disclosed, or represents
that its use would not infringe privately owned rights.  Reference herein
to any specific commercial products process, or service by its trade name,
trademark, manufacturer, or otherwise, does not necessarily constitute or
imply its endorsement, recommendation, or favoring by the United States
Government or any agency thereof, or The Regents of the University of
California.  The views and opinions of authors expressed herein do not
necessarily state or reflect those of the United States Government or any
agency thereof, or The Regents of the University of California.
\end{quotation}
\end{scriptsize}

\vskip 2in

\begin{center}
\begin{small}
{\it Lawrence Berkeley Laboratory is an equal opportunity employer.}
\end{small}
\end{center}

\newpage
\renewcommand{\thepage}{\arabic{page}}
\setcounter{page}{1}

\renewcommand{\thefootnote}{\arabic{footnote}}
\setcounter{footnote}{0}

\section{Introduction}
\label{sec:intro}

The Polonyi problem~\cite{PLB131-59,PRD49-779} is one of the most serious
problems in models based on the $N=1$ supergravity~\cite{NPB212-413}. In a
wide class of supergravity models, the Polonyi field $\phi$, which is
a scalar field related to the supersymmetry (SUSY) breaking,
has a mass $m_\phi$ of the order of the gravitino mass. During
inflation, $\phi$ takes an amplitude of the order of the gravitational
scale $M\equiv M_{\rm pl}/\sqrt{8\pi}\simeq 2.4\times
10^{18}\GEV$. After the inflation, the condensation of $\phi$ starts to
oscillate when the expansion rate of the universe, $H$, becomes
comparable to $m_\phi$ and finally decays into particles in the
observable sector. Since the interactions of the Polonyi field are
suppressed by powers of $M^{-1}$, the decay rate of the Polonyi field,
$\Gamma_\phi$, is very small as
\begin{equation}
    \Gamma_\phi \sim N\frac{m_\phi^3}{M_{\rm pl}^2},
    \label{decay_rate}
\end{equation}
where $N$ is the number of the decay modes. (In the following
calculations, we take $N=100$.)  Therefore, the Polonyi field is
expected to decay when the temperature of the universe becomes very
low. The reheating temperature $T_R$ due to the decay of the Polonyi
field is given by
\begin{equation}
    \label{rtemp}
    T_R \sim 1 {\rm MeV} 
    \left(\frac{m_{\phi}}{10{\rm TeV}}\right)^{3/2}.
\end{equation}
This causes serious cosmological difficulties; namely the Polonyi
field may destroy the great success of the big-bang nucleosynthesis
(BBN), and the entropy production due to its decay may dilute the
baryon number of the universe too much.

In the previous works~\cite{PLB174-176,PLB342-105} it has been pointed
out that the Polonyi problem can be solved if the gravitino mass
$m_{3/2}$ (which is the same order of the Polonyi mass) is larger than
$O(10)\TEV$ in order to hasten the decay of the Polonyi field. Thus,
it is desirable to raise the gravitino mass while keeping the
effective SUSY breaking scale in the observable sector at
$O(100)\GEV$. In no-scale type supergravity models, such mass
hierarchy is realized~\cite{PRD50-2356} and hence the no-scale type
supergravity model with $m_{3/2}\gsim O(10)\TEV$ has been suggested as
an attractive solution to the Polonyi problem.

In reference~\cite{PLB342-105}, however, it has been also pointed out
that the mass density of the lightest superparticle (LSP) produced by
the decay of the Polonyi field may overclose our universe if LSP is
stable. As we will see below, the mass density of LSP increases as the
reheating temperature $T_R$ due to the decay of the Polonyi field
decreases.  Therefore, a lowerbound on $T_R$ is derived requiring that
the present mass density of LSP should not exceed the critical density
of the universe $\rho_c$. In this letter, we obtain the lowerbound on
$T_R$ in the framework of the minimal SUSY SU(5) model with no-scale
type boundary conditions on the SUSY breaking parameters.

\section{The Model}
\label{sec:model}

Before starting cosmological arguments, let us first describe our
basic assumptions. We consider the minimal SUSY SU(5) model with
no-scale type boundary conditions. This model has three types of Higgs
field; $H({\bf 5})$ and $\bar{H}({\bf 5^*})$ which contain flavor
Higgses $H_f$ and $\bar{H}_f$, and $\Sigma ({\bf 24})$ whose
condensation breaks the SU(5) group into the gauge group of the
minimal SUSY standard model (MSSM), $\rm SU(3)_C\times SU(2)_L\times
U(1)_Y$. For the Higgs sector, the superpotential is given by
\begin{equation}
    W = \frac{1}{3} \lambda {\rm tr} \Sigma^3 
    + \frac{1}{2} M_\Sigma {\rm tr} \Sigma^2
    + \kappa \bar{H} \Sigma H
    + M_H \bar{H} H,
    \label{W_su5}
\end{equation}
where $\lambda$ and $\kappa$ are dimensionless constants, while
$M_\Sigma$ and $M_H$ are mass parameters which are of the order of the
grand unified theory (GUT) scale $M_{\rm GUT} (\sim
10^{16}\GEV)$. Furthermore, the model also has the soft SUSY breaking
terms;
\begin{equation}
    {\cal L}_{\rm soft} = 
      - \frac{1}{3} \lambda A_\Sigma {\rm tr} \Sigma^3 
      - \frac{1}{2} M_\Sigma B_\Sigma {\rm tr} \Sigma^2
      - \kappa A_H \bar{H} \Sigma H
      - M_H B_H \bar{H} H +h.c.,
    \label{L_soft_su5}
\end{equation}
where $A_\Sigma$, $B_\Sigma$, $A_H$ and $B_H$ are SUSY breaking
parameters.  Minimising the Higgs potential, we find the following
stationary point;
\begin{equation}
    \vev{\Sigma} = 
    \frac{1}{\lambda} \lmp 
    M_\Sigma + 2 \lsp A_\Sigma - B_\Sigma \rsp
    + O\lsp \frac{A_\Sigma}{M_\Sigma}, \frac{B_\Sigma}{M_\Sigma} \rsp 
    \rmp \times {\rm diag}(2,2,2,-3,-3),
    \label{vac_su5}
\end{equation}
where the SU(5) is broken down to $\rm SU(3)_C\times SU(2)_L\times
U(1)_Y$. Regarding this stationary point as the vacuum, we obtain MSSM
as the effective theory below the GUT scale $M_{\rm GUT}$. Here, the
masslessness of the flavor Higgses $H_f$ and $\bar{H}_f$ is achieved
by a fine tuning among several parameters; $M_H - 3 \kappa
M_\Sigma/\lambda \simeq \mu_H$, where $\mu_H$ is the SUSY-invariant
Higgs mass in MSSM.

In the present model, the parameters in MSSM at the electroweak scale
is obtained by solving renormalization group equations (RGEs). Our
method is as follows. The boundary conditions on the parameters in the
minimal SUSY SU(5) model are given at the gravitational scale $M$.
Since we assume the no-scale type supergravity models, all the SUSY
breaking parameters except for the gaugino mass vanish at the
gravitational scale. From the gravitational scale to the GUT scale,
the parameters follow the renormalization group flow derived from RGEs
in the minimal SUSY SU(5) model. Then we determine the parameters in
MSSM at the GUT scale through an appropriate matching condition
between the parameters in the SUSY SU(5) model and those in MSSM.
Finally, we use RGEs in MSSM from the GUT scale to the electroweak
scale in order to obtain the low energy parameters.

As for the matching condition, we have a comment. In the stationary
point (\ref{vac_su5}), the mixing soft mass term of the two flavor
Higgs bosons, $m_{12}^2\bar{H}_fH_f$, is generated at the tree level,
where $m_{12}^2$ is given by
\begin{equation}
    m_{12}^2 (M_{\rm GUT}) \simeq 
      \llp \frac{6\kappa}{\lambda} 
      (A_\Sigma - B_\Sigma)(A_H - B_\Sigma) 
      -\mu_H B_H.
      \rlp_{\mu =M_{\rm GUT}}
      \label{m12^2}
\end{equation}
Since the mixing mass term depends on unknown parameters, $\lambda$
and $\kappa$ in eq.(\ref{W_su5}), we regard $m_{12}^2$ as a free
parameter taking account of the uncertainty of $\lambda$ and $\kappa$
in our analysis.  Then, the low energy parameters are essentially
determined by the gauge and Yukawa coupling constants and the
following three parameters; the supersymmetric Higgs mass $\mu_H$, the
mixing mass of the two flavor Higgs bosons $m_{12}^2$, and the unified
gaugino mass.\footnote
{In fact, parameters in MSSM slightly depend on the parameters in the
SUSY GUT such as $\lambda$, $\kappa$ and so on. In our numerical
calculation, we ignore the effects of these parameters on the 
renormalization group flow.}
However, it is more convenient to express these parameters by other
physical ones. In fact, one combination of them is constrained so that
the flavor Higgs bosons have correct vacuum expectation values;
$\langle H_f\rangle^2+\langle\bar{H}_f\rangle^2\simeq (174\GEV)^2$. As
the other two physical parameters, we use the mass of LSP, $m_{\rm
LSP}$, and the vacuum angle
$\tan\beta\equiv\langle{H_f}\rangle/\langle{\bar{H}_f}\rangle$. Thus,
once we fix $m_{\rm LSP}$ and $\tan\beta$, we can determine all the
parameters in MSSM.\footnote
{Yukawa coupling constants are determined so that the fermions have
correct masses. The gauge coupling constants are also fixed so that
their correct values at the electroweak scale are reproduced.}

Following the above procedure, we solve the RGEs numerically, and
obtain the low energy parameters in MSSM. Then, we determine the mass
spectrum and the mixing angles for all superparticles. One remarkable
thing is that {\it LSP almost consists of bino $\tilde{B}$ which is
the superpartner of the gauge field for $U(1)_Y$} if we require that
LSP is neutral. Therefore, in our model, the LSP mass $m_{\rm LSP}$ is
essentially equivalent to the bino mass. This fact simplifies the
following analysis very much.

\section{Density of LSP}
\label{sec:density}

Now we are in a position to discuss the mass density of LSP produced
by the decay of the Polonyi field. The decay of the Polonyi field
produces a large number of superparticles, which promptly decay into
LSPs. The number density of LSP produced by the decay, $n_{{\rm
LSP},i}$, is of the same order of that of the Polonyi field
$n_\phi\equiv\rho_\phi /m_\phi$ (with $\rho_\phi$ being the energy
density of the Polonyi field). Just after the decay of the Polonyi
field, the yield variable for LSP, $Y_{\rm LSP}$, which is defined by
the ratio of the number density of LSP to the entropy density $s$, is
given by
\begin{eqnarray}
    m_{\rm LSP} Y_{\rm LSP} & \simeq & \frac{\rho_\phi}{s}
    \simeq \frac{m_{\rm LSP}\rho_{{\rm LSP},i}}{m_\phi s}
    \sim \left(\frac{m_{\rm LSP}T_R}{m_\phi}\right)
    \nonumber \\
    & \sim & 10^{-5}\GEV \left(\frac{m_{\rm LSP}}{100\GEV}\right)
    \left(\frac{T_R}{1\MEV}\right)\left(\frac{10\TEV}{m_\phi}\right),
    \label{mY_LSP}
\end{eqnarray}
where $\rho_{{\rm LSP},i}$ is the mass density of LSP just after the
decay of the Polonyi field. If LSP is stable and the pair annihilation
of LSP is not effective, $Y_{\rm LSP}$ is conserved until today.
Comparing the ratio given in eq.(\ref{mY_LSP}) with the ratio of the
critical density $\rho_c$ to the present entropy density $s_{0}$,
\begin{equation}
    \frac{\rho_c}{s_{0}} \simeq 3.6 \times 10^{-9}h^2~\GEV,
    \label{critical}
\end{equation}
where $h$ is the Hubble constant in units of 100km/sec/Mpc, we see
that LSP overcloses the universe in the wide parameter region for
$m_{\rm LSP}, m_{\phi}$ and $T_R$ which we are concerned with.

If the pair annihilation of LSP takes place effectively, its abundance
is reduced to 
\begin{equation}
    \frac{n_{\rm LSP}}{s} \simeq 
    \left. \frac{H}{s\langle\sigma_{\rm ann}v_{\rm rel}\rangle} 
      \rabs_{T=T_R},
      \label{abundance_LSP}
\end{equation}
where $\sigma_{\rm ann}$ is the annihilation cross section, $v_{\rm
rel}$ is the relative velocity, and $\langle\cdots\rangle$ represents
the average over the phase space distribution of LSP. From
eq.(\ref{abundance_LSP}), we obtain a lowerbound on the annihilation
cross section,
\begin{equation}
    \langle\sigma_{\rm ann}v_{\rm rel}\rangle \gsim
    3\times 10^{-8}h^{-2}\GEV^{-2}
    \left( \frac{m_{\rm LSP}}{100\GEV}\right)
    \left( \frac{100\MEV}{T_R}\right),
    \label{sv_limit}
\end{equation}
in order that the mass density of LSP does not overclose the universe. 

Comparing this bound with the annihilation cross section of LSP, we
derive a bound on the reheating temperature by the decay of the
Polonyi field. Since LSP is most dominated by bino, it annihilates
into fermion pairs. The annihilation cross section is given
by~\cite{PLB230-78}
\begin{equation}
    \langle\sigma_{\rm ann}v_{\rm rel}\rangle 
    = a + b\langle v^2\rangle,
    \label{sigma*v}
\end{equation}
where $\langle v^2\rangle$ is the average velocity of LSP,
and
\begin{eqnarray}
    a  & \simeq &
    \frac{32\pi\alpha_1^2}{27} 
    \frac{m_t^2}{(m_{\tilde{t}_R}^2 + m_{\rm LSP}^2 - m_t^2)^2}
    \lsp 1 - \frac{m_t^2}{m_{\rm LSP}^2} \rsp^{1/2} 
    \theta (m_{\rm LSP}-m_t),
    \label{s-wave} \\
    b &\simeq& \frac{8\pi\alpha_1^2}{3} \sum_{m_f\leq T} Y_f^4 \lmp
    \frac{m_{\rm LSP}^2}{(m_{\rm LSP}^2+m_{\tilde{f}}^2)^2}
    - \frac{2m_{\rm LSP}^4}{(m_{\rm LSP}^2+m_{\tilde{f}}^2)^3}
    + \frac{2m_{\rm LSP}^6}{(m_{\rm LSP}^2+m_{\tilde{f}}^2)^4} \rmp.
    \label{p-wave}
\end{eqnarray}
Here, $\alpha_1^2\equiv g_1^2/4\pi\simeq 0.01$ represents the fine
structure constant for U(1)$_{\rm Y}$, $m_t$ the top-quark mass, $Y_f$
the hypercharge of the fermion $f$, and $m_{\tilde{f}}$ the mass of
the sfermion $\tilde{f}$.  Notice that $a$ and $b$ terms correspond to
$s$- and $p$-wave contributions, respectively. Taking
$m_{\tilde{f}}\sim m_{\rm LSP}\sim 100\GEV$, the annihilation cross
section given in eq.(\ref{sigma*v}) is at most $3\times
10^{-8}\GEV^{-2}$. Using this result in the inequality
(\ref{sv_limit}), we can see that the reheating temperature must be
higher than about 100MeV even if $\langle v^2\rangle\sim 1$. If the
average velocity is smaller than 1, the constraint becomes more
stringent, as we will see below.

\section{Thermalization of LSP}
\label{sec:thermalization}

In order to obtain the precise lowerbound on the reheating temperature
$T_R$, we have to know $\langle v^2\rangle$, as well as the mass
spectrum of the superparticles on which the annihilation cross section
depends. First, let us discuss the averaged velocity of LSP, $\langle
v^2\rangle$. Since LSP is mostly the bino, it loses its energy by
scattering off the background fermions. In the model with the no-scale
type boundary conditions, right-handed sleptons become the lightest
among the sfermions, and hence LSP loses its energy mainly by
scattering off the background electron (and $\mu$ and $\tau$, if the
temperature is higher than their masses). If LSP is relativistic, the
cross section for this process, $\sigma_{\rm scatt}$, is estimated as
\begin{equation}
    \langle \sigma_{\rm scatt} v_{\rm rel} \rangle 
    \simeq 128\pi \alpha_1^2
    \frac{E_{\rm LSP}^2 T_R^2}{m_{\tilde{e}_R}^4 m_{\rm LSP}^2},
    \label{sigma_scatt}
\end{equation}
where $E_{\rm LSP}$ is the energy of LSP, and $m_{\tilde{e}_R}$ the
mass of the right-handed selectron.\footnote%
{This cross section is applied for $E_{\rm LSP} T_{R} \ll
m_{\tilde{e}_R}^2$.}
The energy loss rate for
the relativistic LSP, $\Gamma_{\rm scatt}^{\rm R}$, is given by
\begin{equation}
    \label{loss-rate}
    \Gamma_{\rm scatt}^{\rm R} \simeq 
     n_e \langle \sigma_{\rm scatt} v_{\rm rel} \rangle 
     \frac{\Delta E_{\rm LSP}}{E_{\rm LSP}},
\end{equation}
where $n_e$ represents the number density of the background electron
and $\Delta E_{\rm LSP}$ is the averaged  energy loss of LSP in 
one scattering which is  given by
\begin{equation}
    \label{energy-loss}
    \Delta E_{\rm LSP} \simeq  
    -12E_{\rm LSP}\left(\frac{T_R E_{\rm LSP}}{m_{\rm LSP}^2}\right).
\end{equation}
Taking the ratio of the energy loss  rate $\Gamma_{\rm
scatt}^{\rm R}$ to the expansion rate $H$ of the universe, we find
\begin{equation}
    \left. \frac{\Gamma_{\rm scatt}^{\rm R}}{H}
      \rabs_{E_{\rm LSP}\gg m_{\rm LSP}}
      \simeq 
      2\times10^3 \left( \frac{E_{\rm LSP}}{10^2\GEV} \right)^3
      \left( \frac{T_R}{100\MEV} \right)^4
      \left( \frac{100\GEV}{m_{\tilde{e}_R}} \right)^4
      \left( \frac{100\GEV}{m_{\rm LSP}} \right)^4.
\end{equation}
Thus, if $T_R \gsim$ a few $\times$ 10MeV, the energetic LSP loses its
energy through the scattering off thermal electrons efficiently for
$m_{\tilde{e}_R}\sim m_{\rm LSP}\simeq O(100/GEV)$, and becomes a
non-relativistic particle.

The non-relativistic LSP further loses its energy by scattering off
background electrons. The averaged loss of the kinetic energy for the
non-relativistic LSP in one scattering process, $\Delta \epsilon_{\rm
LSP}$, is given
by\footnote{%
Naively, it is expected that $\Delta \epsilon_{\rm LSP}$ is $\sim
T_R$. However this order of the energy loss is cancelled out when the
average is taken over angles of the incident particles and  the actual
energy loss is much smaller than the naive expectation.}
\begin{equation}
    \Delta \epsilon_{\rm LSP} \simeq
    -\frac{20\epsilon_{\rm LSP} T_R}{m_{\rm LSP}}
    \left( 1 -\frac{T_R}{\epsilon_{\rm LSP}}\right),
    \label{Delta_E}
\end{equation}
where $\epsilon_{\rm LSP}\equiv E_{\rm LSP}-m_{\rm LSP}$ is the
kinetic energy of LSP. As one can see in eq.(\ref{Delta_E}), the LSP
which has a kinetic energy larger than $\sim T_R$ tends to lose its
energy through the scattering process, while LSP receives energy from
the thermal bath if its energy is smaller than $\sim T_R$. Thus, if
the scattering processes take place effectively, the averaged kinetic
energy of LSP becomes $\sim T_R$, {\it i.e.} LSP goes into the
kinetic equilibrium.

The energy loss rate $\Gamma_{\rm scatt}^{\rm NR}$ for the 
non-relativistic LSP is given by
\begin{equation}
    \Gamma_{\rm scatt}^{\rm NR} \simeq 
    n_e \langle\sigma_{\rm scatt} v_{\rm rel}\rangle \times
    \frac{20T_R}{m_{\rm LSP}}
    \simeq 
    \frac{5760\alpha_1^2}{\pi}
    \frac{T_R^6}{m_{\rm LSP}m_{\tilde{e}_R}^4}.
    \label{gamma_scatt}
\end{equation}
The LSP goes into the kinetic equilibrium if the scattering rate
$\Gamma_{\rm scatt}^{\rm NR}$ is larger than the expansion rate of the
universe. Taking $m_{\tilde{e}_R}\sim m_{\rm LSP}$, the ratio of
$\Gamma_{\rm scatt}^{\rm NR}$ to the expansion rate of the universe,
$H$, is given by
\begin{equation}
    \left. \frac{\Gamma_{\rm scatt}^{\rm NR}}{H} \rabs_{E_{\rm LSP}
      \sim m_{\rm LSP}}
      \simeq 4 \times 10^{3} ~
      \lsp \frac{m_{\rm LSP}}{100\GEV} \rsp^{-5}
      \lsp \frac{T_R}{100\MEV} \rsp^{4}.
\end{equation}
Thus, if the reheating temperature is higher than about 10MeV,
produced LSPs go into kinetic equilibrium as far as $m_{\rm LSP} \sim
O(100)$GeV.  Furthermore, as we discussed in the previous section,
the reheating temperature should be higher than at least 100MeV in
order to decrease the mass density of LSP sufficiently. Thus, we
conclude that the produced LSPs go into kinetic equilibrium if we
require that the mass density of the relic LSP should not overclose
the universe.\footnote
{By the numerical calculation we have checked, in fact, that the
scattering rate given in eq.(\ref{gamma_scatt}) is always larger than
the expansion rate of the universe when the relic LSP does not
overclose the universe. (See fig.2.)}
In this case, the averaged velocity is given by
\begin{equation}
    \langle v^2\rangle \simeq \frac{3T_R}{m_{\rm LSP}}.
\end{equation}
>From this we easily see that the LSP abundance given in
eq.(\ref{abundance_LSP}) decreases as the reheating temperature gets
higher. Thus, we obtain the lowerbound on the reheating temperature.

\section{Results}
\label{sec:results}

Once we know the averaged velocity $\langle v^2\rangle$, we can
calculate the annihilation cross section of LSP, and get the
lowerbound on the reheating temperature after the decay of the Polonyi
field. In this letter, we first solve RGEs based on the minimal SU(5)
model with the no-scale boundary conditions, and determine the mass
spectrum of the superparticles. We only investigate the parameter
space which is not excluded by the experimental or theoretical
constraints. The constraints which we use are as follows:
\begin{itemize}
\item Higgs bosons $H_f$ and $\bar{H}_f$ have correct vacuum
expectation values.
\item Perturbative picture is valid below the gravitational scale.
\item LSP is neutral.
\item Sfermions (especially, charged sleptons) have masses larger than
the experimental lower limits~\cite{PDG}.
\item The branching ratio for $Z$-boson decaying into neutralinos is 
not too large~\cite{PLB350-109}.
\end{itemize}
Then, with the obtained mass spectrum of superparticles, we calculate
the annihilation cross section and determine the lowerbound on the
reheating temperature from the following equation;
\begin{equation}
    \left. \frac{H}{s\langle\sigma_{\rm ann}v_{\rm rel}\rangle} 
      \rabs_{T=T_R} \leq
      \frac{\rho_c}{s_0} \simeq 3.6h^2 \times 10^{-9}\GEV.
\end{equation}

In fig.~1, we show the lowerbound on the reheating temperature in the
$\tan\beta$ vs. $m_{\rm LSP}$ plane. In the figures, large or small
$\tan\beta$'s are not allowed since the Yukawa coupling constant for
the top quark or bottom quark blows up below the gravitational scale
for such $\tan\beta$'s. Furthermore, there also exists a lowerbound on
the LSP mass. In the case where $\tan\beta\lsim 20$, charged sfermions
become lighter than the experimental limit if the LSP mass becomes
lighter than $\sim 50\GEV$.  On the other hand, for the large
$\tan\beta$ case, unless the bino mass is sufficiently large, the
lightest charged slepton becomes LSP. (Remember that the dominant
component of LSP is bino.)  Thus, the lowerbound on $m_{\rm LSP}$ is
obtained. As we can see, the reheating temperature should be larger
than about 100MeV, even for the case where $m_{\rm LSP}\sim 50\GEV$.
The constraint becomes more stringent as $m_{\rm LSP}$ increases,
since the masses of the superparticles which mediate the annihilation
of LSP becomes larger as the LSP mass increases.

If we translate the lowerbound on the reheating temperature into that
of the Polonyi mass $m_\phi$, we obtain $m_\phi\gsim 100\TEV$ (see
eq.(\ref{rtemp})). We can also see that the lowerbound is almost
independent of $\tan\beta$. In fig.~2, We show the lowerbound on $T_R$
as a function of the LSP mass for $\tan\beta =10$, and $\mu_H >0$.

Here, we should comment on the accidental case where the annihilation
process hits the Higgs pole in the $s$-channel. If the LSP mass is
just half of the lightest Higgs boson mass, the LSP annihilation cross
section is enhanced since LSP has small but nonvanishing fraction of
higgsino component. If the parameters are well tuned, such a situation
can be realized and the lowerbound of $T_R$ decreases to $O(10)\MEV$.
However, we consider that such a scenario are very unnatural since a
precise adjustment of the
parameters is required in order to hit the Higgs pole.\footnote
{In the case where the annihilation process hits the pole of heavier
Higgs bosons, the cross section is not enhanced so much, since the
widths of the heavier Higgs bosons are quite large.}

\section{Conclusions}
\label{sec:conclusions}

In this letter, we have obtained the lowerbound on the reheating
temperature due to the decay of the Polonyi field in a framework of
the no-scale type supergravity model. As a result, we have seen that
the Polonyi mass should be larger than about 100TeV which may raise a
new fine-tuning problem~\cite{PLB173-303}.

We have assumed the minimal SUSY GUT model in the present
analysis. However, the main conclusion is not changed as far as LSP
is mostly the bino, because the minimum value of the lowerbound ($T_R
\simeq 100$MeV) is obtained when the mass of the selectron takes the
experimentally allowed lower limit.

We have assumed that LSP is stable so far. However, if we introduce
$R$-parity violation, LSP becomes unstable and the allowed $T_R$ is as
low as a few MeV. It is also the case if we assume a very light LSP
such as a neutral higgsino~\cite{PLB131-59} or an
axino~\cite{NPB358-447} whose masses are less than about 100MeV.

\section*{Acknowledgement}

Two of the authors (T.M. and T.Y.) would like to thank M.~Yamaguchi for
useful discussions in the early stage of this work.

\newpage
%
%
\newcommand{\Journal}[4]{{\sl #1} {\bf #2} {(#3)} {#4}}
\newcommand{\APJ}{Ap. J.}
\newcommand{\CJP}{Can. J. Phys.}
\newcommand{\NC}{Nuovo Cimento}
\newcommand{\NP}{Nucl. Phys.}
\newcommand{\PL}{Phys. Lett.}
\newcommand{\PR}{Phys. Rev.}
\newcommand{\PRep}{Phys. Rep.}
\newcommand{\PRL}{Phys. Rev. Lett.}
\newcommand{\PTP}{Prog. Theor. Phys.}
\newcommand{\SJNP}{Sov. J. Nucl. Phys.}
\newcommand{\ZP}{Z. Phys.}

\newpage

\noindent
{\large\bf Figure Caption} 

\vskip 0.5cm

\noindent
{\bf fig. 1}

\noindent 
Lowerbound on $T_R$ is shown in $\tan\beta$ vs. $m_{\rm LSP}$
plane. The meaning of each marks is as follows; $\circ : 100\MEV\leq
T_R\leq 500\MEV$, $\times : 500\MEV\leq T_R\leq 1\GEV$, $\Box :
1\GEV\leq T_R\leq 5\GEV$, $+ : 5\GEV\leq T_R\leq 10\GEV$, $\Diamond :
10\GEV\leq T_R\leq 50\GEV$. The sign of the SUSY-invariant Higgs mass
$\mu_H$ is taken to be (a) $\mu_H >0$, and (b) $\mu_H <0$.

\noindent
{\bf fig. 2}

\noindent 
Lowerbound on $T_R$ is shown as a function of $m_{\rm LSP}$. Here, we 
take $\tan\beta =10$ and $\mu_H >0$ in the solid line. Furthermore, the 
temperature at which the relation $\Gamma_{\rm scatt}=H$ realizes is 
also shown in dashed line.

\newpage

\begin{figure}[p]
\epsfxsize=13cm
\centerline{\epsfbox{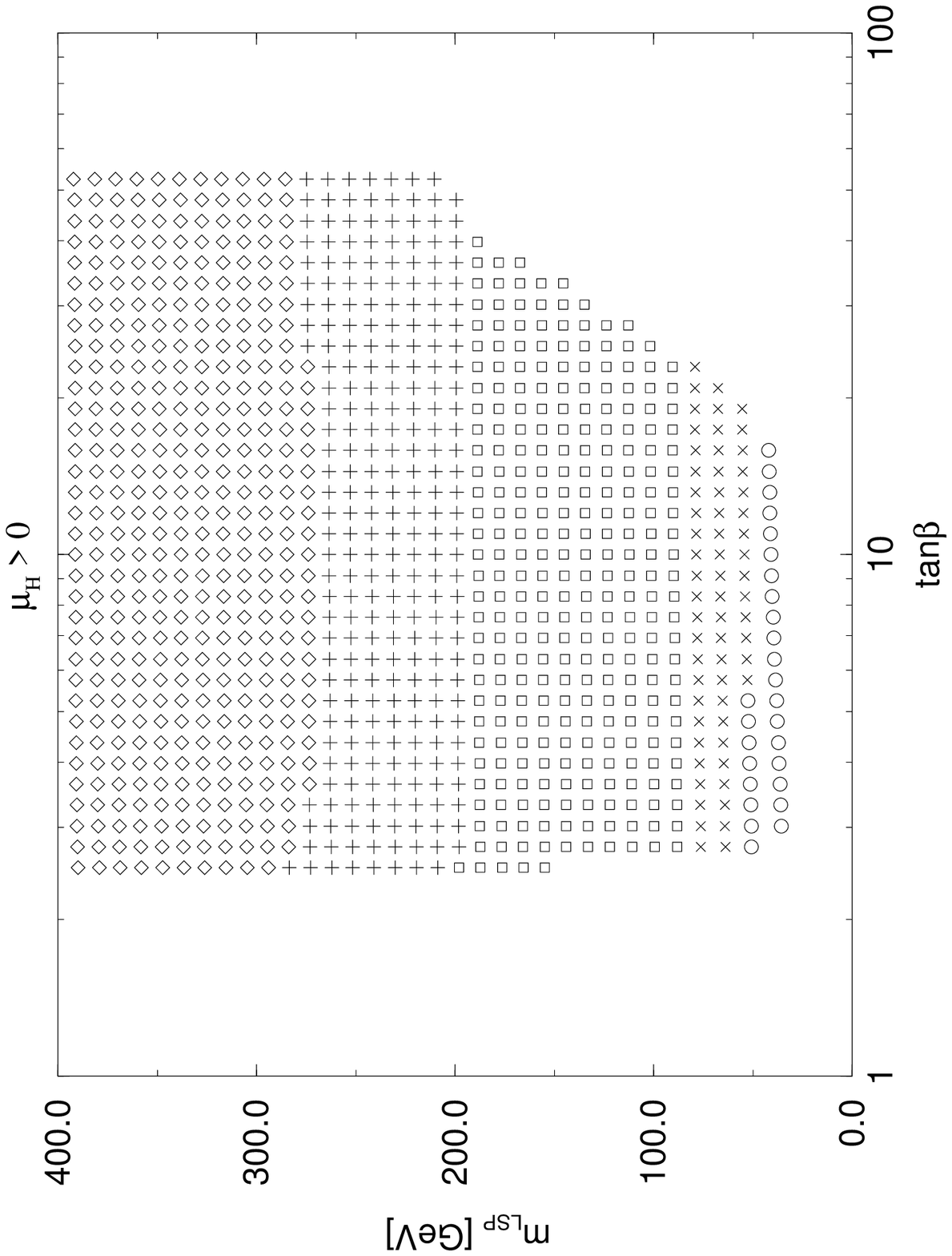}}
\vskip .5in
\begin{center}
{\bf \large fig. 1a}
\end{center}
\end{figure}

\begin{figure}[p]
\epsfxsize=13cm
\centerline{\epsfbox{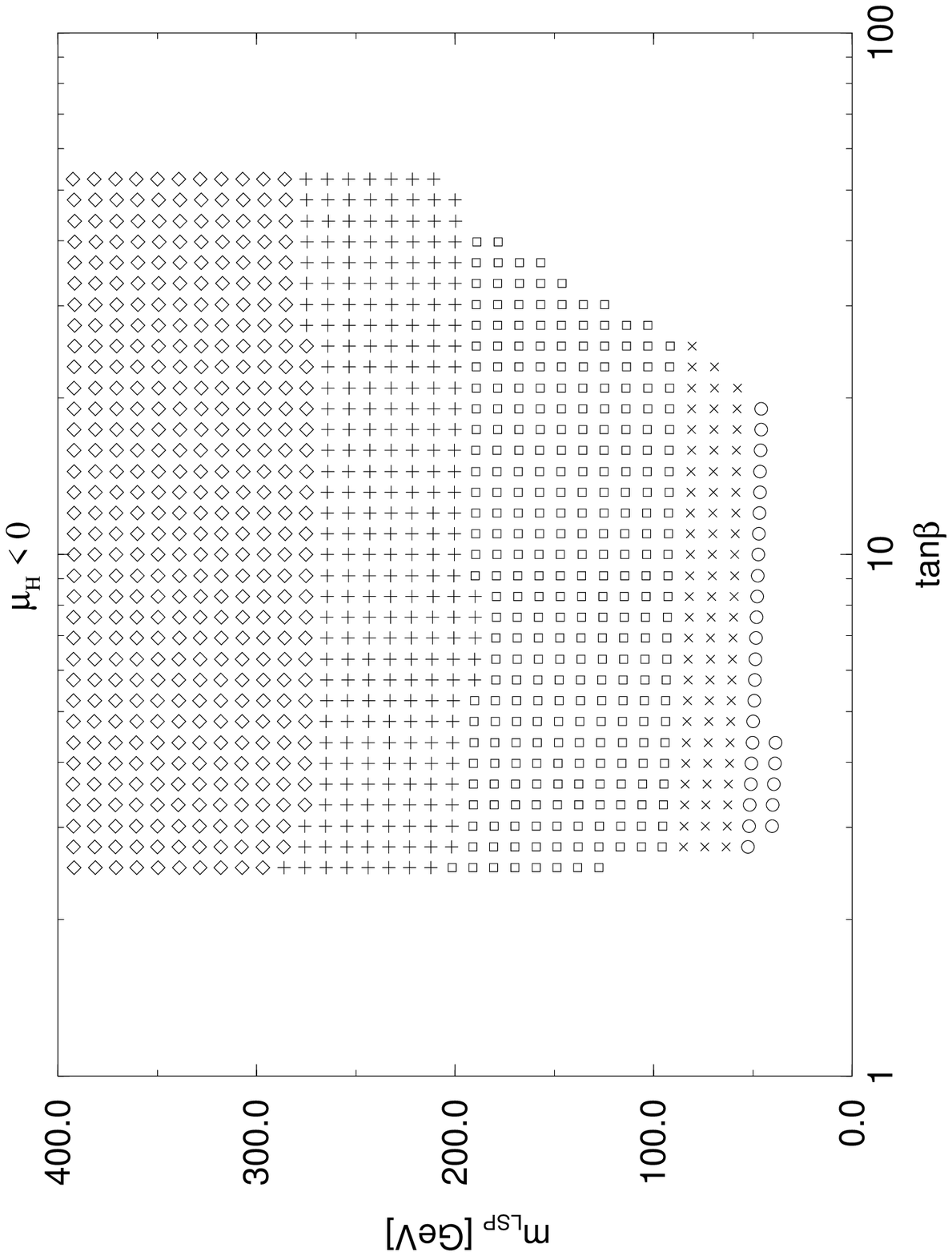}}
\vskip .5in
\begin{center}
{\bf \large fig. 1b}
\end{center}
\end{figure}

\begin{figure}[p]
\epsfxsize=13cm
\centerline{\epsfbox{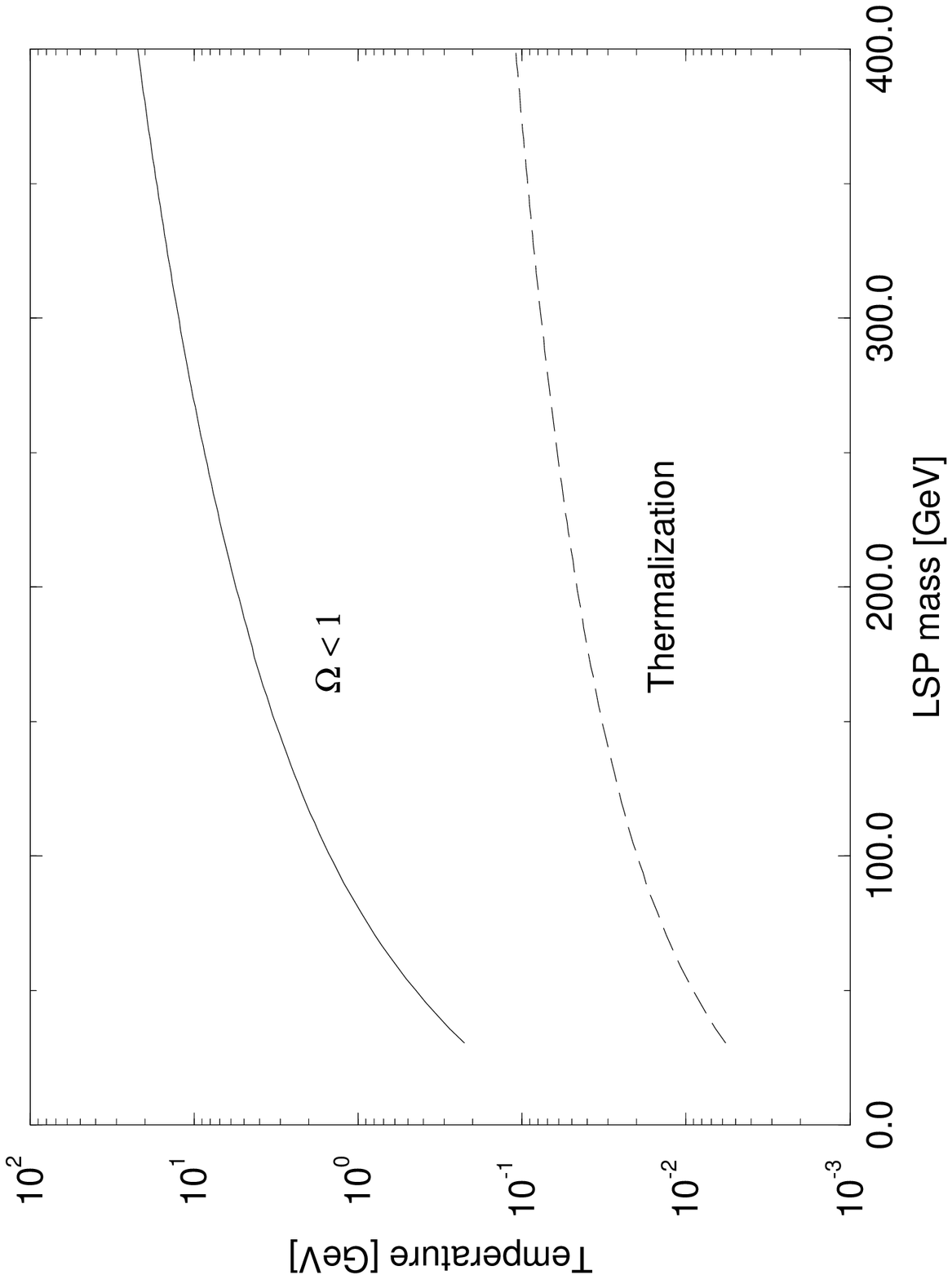}}
\vskip .5in
\begin{center}
{\bf \large fig. 2}
\end{center}
\end{figure}

\end{document}